\documentclass[10pt,a4paper]{article}

\usepackage{times}

\usepackage{epsfig,wrapfig}
\usepackage{epstopdf} % for pdflatex

\usepackage{fancyhdr}

\usepackage[lmargin=2.5cm, rmargin=2.5cm,tmargin=3.50cm,bmargin=3.50cm]{geometry}
\usepackage{indentfirst}

\usepackage{titlesec}
\titleformat{\section}[hang]
  {\centering}{\thesection}{1ex}{\normalsize \textsc}%%
\titleformat{\subsection}[hang]
  {}{\thesubsection}{1ex}{\normalsize \textit}%%

\newcommand{\acknowledgement}{\section*{\centering{\textnormal{\normalsize{\textsc{Acknowledgement}}}}}}

\renewcommand{\thesection}{ \normalsize \textnormal{\Roman{section}.}}
\renewcommand{\thesubsection}{\normalsize \textnormal{\textsc{\textit{\Alph{subsection}.}}}}

\def\e{\begin{equation}}
\def\f{\end{equation}}
\def\_#1{{\bf #1}}

\def\.{\cdot}

\usepackage{color,soul} % for yellow-highlighting text
\begin{document}

%%% Title of paper
\title{\large \textbf{
Analysis and Design of a Reconfigurable Metasurface based on Chalcogenide Phase-Change Material for Operation in the Near and Mid Infrared
}}
%
%%% Author(s) and affiliation
\def\affil#1{\begin{itemize} \item[] #1 \end{itemize}}
\author{\normalsize \bfseries
Alexandros Pitilakis$^{1,2}$, {Alexandros Katsios}$^{1}$ and
Alexandros-Apostolos A. Boulogeorgos$^{1}$
}
\date{}
\maketitle

\vspace{-6ex}
\affil{\begin{center}\normalsize 
$^1$University of Western Macedonia, Electrical and Computer Engineering, 50100, Kozani, Greece\\
$^2$Aristotle University of Thessaloniki, Electrical and Computer Engineering, 54124, Thessaloniki, Greece\\
e-mail: {alexpiti@auth.gr}
\end{center}}

\vspace{-0.5cm}
%%% Abstract
\begin{abstract}
\noindent \normalsize
\textbf{\textit{Abstract} \ \ -- \ \
%%% Start here with text of abstract
We analyze, design and assess the performance of a reflective reconfigurable metasurface (MS) architecture for optical wireless communications. The device is based on the Ge$_2$Sb$_2$Te$_5$ (GST) phase-change material (PCM) alloy, thermally toggled between highly distinct amorphous and crystalline phase-states. We employ simple conductive MS patterns to tune its resonance frequency, while allowing the unit cell response to be analytically predicted using transmission line theory and equivalent circuits. The GST material dispersion is computed in its two extreme phase-states using a Drude/Tauc-Lorentz model (DTLM), whose parameters are fitted to state-of-the-art experimental data; the dispersion in intermediate partially crystallized phase-states is computed using the Lorentz-Lorenz formula. Our results, corroborated by full-wave simulations, demonstrate the potential of PCM materials for the implementation of continuously reconfigurable holographic metasurfaces operating in the infrared bands.}
\end{abstract}

% ---------------------------------------------------------------------
\section{Introduction}
\vspace{-0.2cm}
% ---------------------------------------------------------------------
Reconfigurable metasurfaces have recently emerged as game-changers in wireless communication networks operating in the microwave bands \cite{Pitilakis2018}. Interest in realizing similar functionalities in the optical band has followed \cite{Kim2022}, but reconfigurable optical MS present new and unique challenges. Specifically, electro-optical schemes based on bulk crystals and semiconductors are banded by fractional refractive index changes whereas liquid crystal schemes struggle to reduce the cell size to subwavelength dimensions without sacrificing efficiency. In view of these challenges, chalcogenide PCM technology \cite{Zhang2021Myths} has matured and emerged as a potential enabler, owing to the unity-order change in their refractive index, when transitioning between two highly distinct phase-states; the comparative disadvantages of PCM devices are the high optical losses owing to material absorption and the slow speed and lack of isolation owing to the thermal phase-state transitions and heat diffusion, respectively.

In this work, we employ the most popular PCM for near infrared (NIR) operation, i.e., chalcogenide GST$_{225}$, which is quasi-transparent in its amorphous (room-temperature) state and transitions to a very high index crystallized state when heated; increased absorption in the crystallized state is the main drawback in the NIR, which vanishes in the mid infrared (MIR), $\lambda_0>3~\mu$m. We design a unit cell based on square metallic patches residing on a GST-comprising metal-backed dielectric slab multilayer. Using analytical tools checked against full-wave simulations, we show that controlling the crystallization ratio of the GST slab can widely tune the reflection phase while maintaining high reflection magnitude. Such an architecture can be used in tailored or thermally reconfigurable meta-mirrors, gratings, and phased arrays, effectively allowing for arbitrary wavefront shaping \cite{de2018nonvolatile}.

% ---------------------------------------------------------------------
\section{Methodology}
\vspace{-0.2cm}
% ---------------------------------------------------------------------
\subsection{Evaluating Reflection Spectra using Transmission Line and Equivalent Circuit Models}
We select a simple subwavelength unit cell design, depicted in Fig.~\ref{Fig1}(a), consisting of an ultrathin dielectric multilayer lying on top of a metallic ground plane; the multilayer consists of a GST slab sandwiched between two identical silicon dioxide slabs. A simple metallic patterning composed of a square patch resides on the air/oxide interface; the patch size can tune the MS resonance to the desired operating wavelength, its square shape ensures polarization-insensitive operation in a wide cone, and its simple form factor allows the use of well-established equivalent circuit models (ECM) to accurately evaluate its dispersion near the main resonance for oblique incidence in both principal polarizations \cite{Luukkonen2008}. The complex-valued reflection coefficient spectra of this unit cell can be analytically computed using transmission line modeling (TLM), i.e., ABCD matrices. This TLM/ECM formulation can be used in two ways: to compute the response (such as the phase-slope and maximum absorption) for given cell parameters (e.g., thicknesses, widths, materials) or to optimize the cell parameters for a target response. Coupled-mode theory can be alternatively employed to reveal the importance of the balance between the dissipative and radiative properties of the unit cell modeled as an open resonator \cite{park2020an}. 

Of note, in this study, all metallic parts are assumed to be perfect electric conductors (PEC) of zero thickness, because the losses in this structure are dominated by GST absorption; in a real-world device the PEC sheets would be replaced by few 10~nm-thick gold, silver, or aluminum films. Furthermore, optically transparent heat-conductive electrodes could be introduced in the multilayer, for the electro-thermal activation of the PCM \cite{de2018nonvolatile}.

\subsection{Computing the Phase-Change Material Frequency Dispersion}
\begin{figure}[]
    \centering 
    \includegraphics[width=130mm]{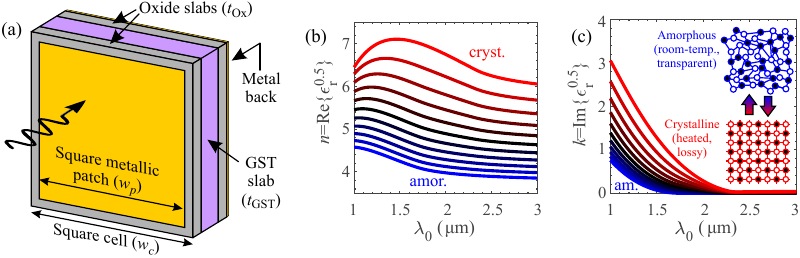}   
    \vspace{-0.3cm}
    \caption{(a) Unit cell architecture. DTLM/LLF-computed refractive index spectra, (b) real and (c) imaginary part.} 
    \label{Fig1}
\end{figure}

The TLM/ECM framework allows the inclusion of material dispersion for all dielectric slabs. We focus on GST dispersion, i.e., on its relative permittivity spectra $\varepsilon_r(\omega) = \varepsilon_1(\omega) -j\varepsilon_2(\omega)$ in a wide band, from the visible to the MIR. Following \cite{Jellison1996}, we first compute the imaginary part $\varepsilon_2(\omega)$ using Drude and Tauc-Lorentz fitting for the low- and high-frequency parts of the spectrum, respectively; the parameters of this combined DTLM for the amorphous and fully crystallized phase-states can be fitted to experimental measurements; in this work, we use the data from \cite{chew2017chalcogenide}. When $\varepsilon_2(\omega)$ is known in a broad spectral band, we compute $\varepsilon_1(\omega)$ using Kramers-Kronig relations, i.e., by $\omega$-integration involving $\varepsilon_2(\omega)$. Finally, the $\varepsilon_r$ spectra for partially crystallized states are computed by the Lorentz-Lorenz formula (LLF) \cite{Chen2015}, and naturally lie between the two extreme spectra. Characteristic NIR spectra for the complex refractive index $n_c=\sqrt{\varepsilon_r}=n-jk$ can be seen in Fig.~\ref{Fig1}(b) and (c), for eleven crystallization ratios from zero (amorphous) to one (crystallized); these have been computed with the DTLM and LLF. These intermediate phase-states can be used for multi-bit (non-binary) digitally-encoded finite-aperture MSs. 

% \begin{equation}
%     E = m c^2
% \end{equation}

%\hl{Show a Table with some reference values at 1550~nm?}

% ---------------------------------------------------------------------
\section{Results}
\begin{figure}[]
    \centering 
    \includegraphics[width=120mm]{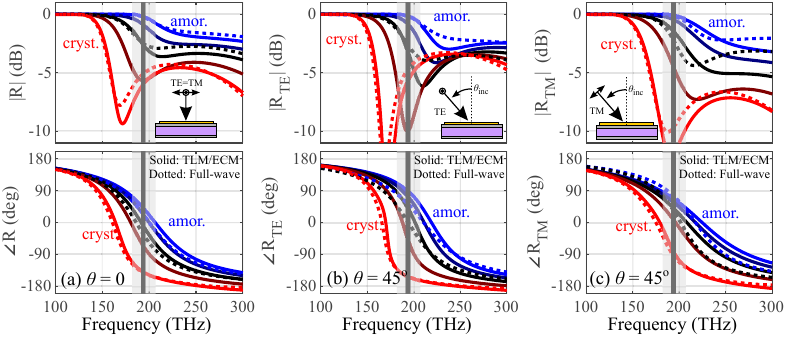}
    \vspace{-0.3cm}
    \caption{TLM/ECM-computed reflection spectra, including GST material dispersion, under (a) normal, (b) oblique TE $45^\circ$, and (c) oblique TM $45^\circ$ incidence, for a range of crystallization ratios. The shaded region marks a 200~nm span centered around the operating wavelength $\lambda_0=1550$~nm (193.4~THz).} 
    \label{Fig2}
\end{figure}

\vspace{-0.2cm}
% ---------------------------------------------------------------------
We evaluate the complex reflection coefficient spectra for normal incidence on the unit cell using the TLM/ECM. We subsequently optimize the oxide and GST slab thicknesses, together with the cell and metal patch widths. The target performance at $\lambda_0=1550$~nm is (i) phase coverage exceeding $180^\circ$ when tuning the GST between its amorphous and crystallized phase-states and (ii) reflection coefficient magnitude larger than $-6$~dB at both states. The TLM/ECM spectra for an indicative design of $w_c=200$~nm, $w_p=0.9w_c$, and $t_\mathrm{Ox}=t_\mathrm{GST}=30$~nm (total thickness is $90$~nm) are depicted in Fig.~\ref{Fig2} with solid curves. While the performance is agreeable in normal incidence, we note the deterioration in oblique incidence; nevertheless, for the TE $45^\circ$ case, a phase span of $180^\circ$ with $-6$~dB magnitude can be found for the pair of 100\% and 60\% crystallization ratios. Finally, we point out that the agreement between TLM/ECM-computed and full-wave simulated spectra (dotted curves in Fig.~\ref{Fig2}, computed with CST Microwave Studio) is acceptable and improves as the multilayer thickness increases, as the index-contrast decreases, and as the elevation angle decreases (towards normal).

{Concerning the electro-thermal aspects, we anticipate that due to the sub-micron lateral dimensions and thickness of the cell, the heating efficiency and speed will be rather high. Moreover, the presence of the ground plane, which acts also as a heat sink, will limit the thermal crosstalk between adjacent cells; preliminary calculations using a lumped equivalent thermal circuit indicate a parasitic temperature increase below 10\% in unheated neighbors. However, full 2D/3D multiphysical simulations are required to properly evaluate the performance and limitations.}

% ---------------------------------------------------------------------
\section{Conclusion and Outlook}
\vspace{-0.2cm}
% ---------------------------------------------------------------------
A simple GST-based reconfigurable unit cell architecture has been designed using efficient analytical tools and evaluated by full-wave simulations. Our results show that PCM-based MSs can enable holographic and wavefront-shaping capabilities for next-generation optical wireless NIR communications NIR. Future research will be devoted to optimizing the thermoelectric aspects of the reconfiguration and accurately evaluating its efficiency and speed.

\acknowledgement
\vspace{-0.2cm}
{This work has received support by the research project MINOAS implemented in the framework of H.F.R.I call ``Basic research Financing (Horizontal support of all Sciences)'' under the National Recovery and Resilience Plan ``Greece 2.0'' funded by the European Union – NextGenerationEU (H.F.R.I. Project Number: 15857). The Authors acknowledge insightful discussions with and computational resources provided by Prof. Emmanouil E. Kriezis.}

%%% References

{\small

%\bibliographystyle{IEEEtran}
%\bibliography{IEEEabrv,RefsList}

\begin{thebibliography}{1}
\providecommand{\url}[1]{#1}
\csname url@samestyle\endcsname
\providecommand{\newblock}{\relax}
\providecommand{\bibinfo}[2]{#2}
\providecommand{\BIBentrySTDinterwordspacing}{\spaceskip=0pt\relax}
\providecommand{\BIBentryALTinterwordstretchfactor}{4}
\providecommand{\BIBentryALTinterwordspacing}{\spaceskip=\fontdimen2\font plus
\BIBentryALTinterwordstretchfactor\fontdimen3\font minus \fontdimen4\font\relax}
\providecommand{\BIBforeignlanguage}[2]{{%
\expandafter\ifx\csname l@#1\endcsname\relax
\typeout{** WARNING: IEEEtran.bst: No hyphenation pattern has been}%
\typeout{** loaded for the language `#1'. Using the pattern for}%
\typeout{** the default language instead.}%
\else
\language=\csname l@#1\endcsname
\fi
#2}}
\providecommand{\BIBdecl}{\relax}
\BIBdecl

\bibitem{Pitilakis2018}
A.~Pitilakis \emph{et al.}, ``Software-defined metasurface paradigm: Concept, challenges, prospects,'' in \emph{2018 12th International Congress on Artificial Materials for Novel Wave Phenomena (Metamaterials)}. IEEE, Aug. 2018, p. 483–485.

\bibitem{Kim2022}
J.~Kim \emph{et al.},``Tunable metasurfaces towards versatile metalenses and metaholograms: A review,'' \emph{Advanced Photonics}, vol.~4, no.~02, Mar. 2022.

\bibitem{Zhang2021Myths}
Y.~Zhang \emph{et al.}, ``Myths and truths about optical phase change materials: A perspective,'' \emph{Applied Physics Letters}, vol. 118, no.~21, May 2021.

\bibitem{de2018nonvolatile}
C.~R. de~Galarreta \emph{et al.}, ``Nonvolatile reconfigurable phase‐change metadevices for beam steering in the near infrared,'' \emph{Advanced Functional Materials}, vol.~28, 1 2018.

\bibitem{Luukkonen2008}
O.~Luukkonen \emph{et al.},``Simple and accurate analytical model of planar grids and high-impedance surfaces comprising metal strips or patches,'' \emph{IEEE Transactions on Antennas and Propagation}, vol.~56, no.~6, p. 1624–1632, Jun. 2008.

\bibitem{park2020an}
J.~Park \emph{et al.}, ``An over‐coupled phase‐change metasurface for efficient reflection phase modulation,'' \emph{Advanced Optical Materials}, vol.~8, 8 2020.

\bibitem{Jellison1996}
G.~E. Jellison and F.~A. Modine, ``Parameterization of the optical functions of amorphous materials in the interband region,'' \emph{Applied Physics Letters}, vol.~69, no.~3, p. 371–373, Jul. 1996.

\bibitem{chew2017chalcogenide}
L.~T. Chew \emph{et al.}, ``Chalcogenide active photonics,'' \emph{Active Photonic Platforms IX}, 8 2017.

\bibitem{Chen2015}
Y.~Chen \emph{et al.}, ``Engineering the phase front of light with phase-change material based planar lenses,'' \emph{Scientific Reports}, vol.~5, no.~1, Mar. 2015.

\end{thebibliography}

}

\end{document}